\documentclass[prc,aps,amsmath,amssymb,nofootinbib,superscriptaddress,12pt]{revtex4-2}
\usepackage{epsfig}
\usepackage[bookmarksnumbered,bookmarksopen,colorlinks,citecolor=blue,linkcolor=blue]{hyperref}
\usepackage{amsmath}
\usepackage{amssymb}
\usepackage{tipa}

\begin{document}

\title{ Spontaneous fission half-lives for heavy and super-heavy nuclei from phenomenological models  }

\author{Yi Xie}
\affiliation{ Department of Physics,
	Guangxi Normal University, Guilin 541004, People's Republic of
	China }

\author{Ning Wang}
\email{wangning@gxnu.edu.cn}
\affiliation{ Department of Physics,
	Guangxi Normal University, Guilin 541004, People's Republic of
	China }
\affiliation{ Guangxi Key Laboratory of Nuclear Physics and Technology, Guilin 541004, People's Republic of
	China }	

\author{Zhongzhou Ren}
\affiliation{School of Physics Science and Engineering, Tongji University, Shanghai 200092, People’s Republic of China}

\begin{abstract}
	A phenomenological model is proposed for a systematic description of the spontaneous fission (SF) half-lives $T_{\rm SF}$ of heavy and super-heavy nuclei. Based on the effective tunneling barrier (ETB), the proposed approach reproduces the SF half-lives of 79 known nuclei with an average deviation of 0.8, which is $17\%$ smaller than that of the linear correlation approach recently proposed in [N. S. Moiseev, N. V. Antonenko and G. G. Adamian, Phys. Rev. C \textbf{112}, 034607 (2025)]. For superheavy nuclei with $45\leqslant N-Z \leqslant 61$, the predicted SF half-lives from these two different phenomenological models are in good agreement with each other. The ETB calculations implies that the $\beta$-decay energy affects the SF half-lives of nuclei far from the $\beta$-stability line. For superheavy nuclei around the magic number $N=184$, the predicted $T_{\rm SF}$ of $^{304}$120 is much shorter than that of $^{298}$Fl. With predicted values of about $10 \sim 160$ ms for $T_{\rm SF}$, the unmeasured SHN $^{293}119 $ could survive for long enough to reach the focal-plane detector in detection systems like the gas-filled recoil separator SHANS in Lanzhou.
	
\end{abstract}

\maketitle

\newpage

\begin{center}
\textbf{I. INTRODUCTION}
\end{center}

As one of the key and sensitive physical inputs, fission barriers and spontaneous fission half-lives of nuclei are frequently used in the studies of nuclear physics \cite{Fan00,Zhao15,Zhang19,Zhang21,Hof16}, reactor physics \cite{Wag77} and nuclear astrophysics \cite{Pan05,Gor15}. Studies of nuclear fission are therefore of great interest \cite{Co63,Bj80,Cap09,And18,Liu96,Pei21,Deni22} and the process of spontaneous fission \cite{Swiat55, Ren05,Ren08,San10,Bao15,Hess,Ren24,Pomo22,Anto25} has been the subject of extensive investigations for nearly eight decades. Accurate prediction of SF half-lives in heavy and superheavy nuclei (SHN) is essential for synthesizing new elements and understanding nuclear structure.
SHN are typically produced via heavy-ion fusion reactions, forming excited compound nuclei that decay via evaporation or fission. For successful identification in detection systems like the gas-filled recoil separator SHANS in Lanzhou \cite{Zhou21,Yang24}, the compound nucleus must survive for longer than $\sim$ 1 $\mu s$ to reach the focal-plane detector. Consequently, accurate predictions of SF half-lives are indispensable for guiding experiments aimed at synthesizing new SHN and characterizing their decay chains.

Compared to $\alpha$-decay, where theoretical models achieve relatively higher reliability \cite{Tian24,Wang15,Royer10,He25}, SF half-life predictions exhibit significant uncertainties, particularly for odd-$A$ and odd-odd nuclei, where deviations from experimental data can span up to five orders of magnitude \cite{Bao15,Hess,Ren24}. This discrepancy stems from the complex, multi-dimensional nature of fission dynamics, involving uncertainties in fragment mass/charge distributions, neutron emission, and energy release. While $\alpha$-decay can be effectively modeled as quantum tunneling through a one-dimensional barrier, fission involves traversing along the complicated potential energy surface (expressed in terms of several deformation parameters, and influenced by nuclear shell effects and pairing correlations) from the ground state to the scission point \cite{Rod14}. To compute deformed mean-field configurations and collective inertias, some microscopic approaches such as the constrained Hartree-Fock-Bogoliubov (HFB) method together with the Gogny energy density functional \cite{Rod20}  and the multidimensional constrained covariant density functional theory \cite{Zhou12} are developed. As a crude approximation, the spontaneous fission half-lives ($T_{\rm SF}$) could be linked to the static fission barrier height $B_{\rm f}$ \cite{Moll15,Wang24}. However, as highlighted by Heßberger, $T_{\rm SF}$ is not uniquely determined by $B_{\rm f}$ alone. For example, the SF half-life of $^{236}$U is up to 9 orders of magnitude longer than that of $^{246}$Cm despite the fission barrier heights \cite{Cap09} of these two nuclei ($B_{\rm f} \approx 6$ MeV) are similar. This suggests additional dynamical factors govern the fission process.

  \begin{figure}
	\setlength{\abovecaptionskip}{ -2.0 cm}
	\includegraphics[angle=0,width=0.85\textwidth]{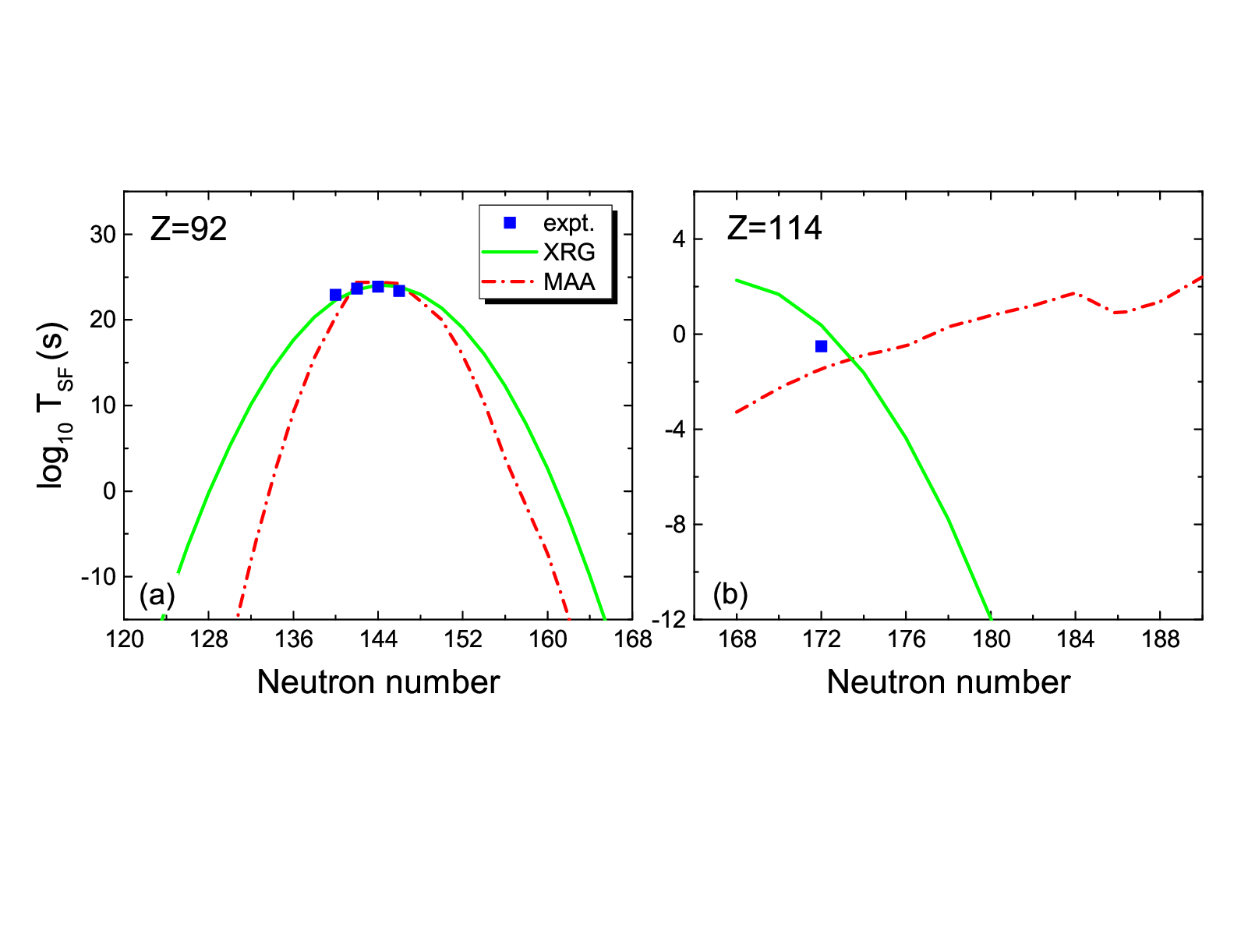}
	\caption{  Comparison of the predicted SF half-lives for even-even Uranium and Flerovium isotopes. The squares denote the experimental data taken from NUBASE2020 \cite{NUBASE2020}. The solid curve and the dot-dashed curve denote the results of XRG formula proposed in \cite{Ren08} and those of MAA formula in \cite{Anto25}, respectively.    }
\end{figure}

Drawing parallels to $\alpha$-decay theory, Xu et al. proposed the relative Coulomb barrier height between the two fission fragments comparing to the $Q$-value in SF as the key quantity for SF tunneling, and the penetration probability $P$ of SF is expressed as \cite{Ren08},  
\begin{eqnarray}
	P =  \exp [-2\pi (V_{\rm top} - Q_{\rm sf})/\hbar \omega_{\rm f} ].
\end{eqnarray}
Here, $V_{\rm top}$ is the height of the Coulomb barrier between fission fragments and $Q_{\rm sf}$ is the SF $Q$-value in the fission process. With an empirical formula for describing the Coulomb barrier,
\begin{eqnarray}
	V_{\rm top} =  a_1 A + a_2 Z^2 + a_3 Z^4 + a_4 (N-Z)^2,
\end{eqnarray}
 the values of $\log_{10} T_{\rm SF}$ for known even-even nuclei can be reproduced reasonably well (with an average deviation of 0.98). In addition, very recently, Moiseev et al. find a linear correlation between $\log_{10} T_{\rm SF}$  and the corresponding $\alpha$-decay energy $Q_\alpha$  for even-even nuclei with the same neutron excess $N-Z$, 
 \begin{eqnarray}
 \log_{10} T_{\rm SF} =  b_0   + b_1 (N-Z) + b_2 (N-Z)^2 + b_3 Q_\alpha,
 \end{eqnarray} 
 with which a phenomenological approach is presented for predicting the spontaneous fission half-lives of actinides and superheavy nuclei. The calculated half-lives closely match the experimental data for known nuclei with an average deviation of 1.0 order of magnitude.

 As a competition between the nuclear force and the Coulomb repulsion, the SF is strongly influenced by the isospin effect and the microscopic structure effects (e.g. shell and pairing effects). In Eq.(2) and Eq.(3), the isospin effect is represented by the neutron excess $(N-Z)$ terms. The microscopic effects are partly considered through $Q_\alpha$ in Eq.(3). For super-heavy nuclei and nuclei far from the $\beta$-stability line, the uncertainties of these two phenomenological formulas are quite large. For example,  Fig. 1  shows the predicted $\log_{10} T_{\rm SF}$ for even-even nuclei with $Z=92$ and $114$ by using Eq.(2) and Eq.(3). One can see that the prediction discrepancies between two models are quite large for unknown nuclei. It is therefore necessary to improve the accuracy of SF half-live predictions for unstable heavy and superheavy nuclei.

In this work, two phenomenological models are used to systematically investigate the trend of the SF half-lives of heavy and super-heavy nuclei, considering that the microscopic calculations are time-consuming. In addition, we also study the influence of the $\beta$-decay energy $Q_\beta$, the $\alpha$-decay energy $Q_\alpha$ and the shell gap $\Delta$ which contains information about the microscopic shell and pairing energies on the SF half-lives of nuclei far from the $\beta$-stability line.

\begin{center}
\textbf{ II. THEORETICAL FRAMEWORKS  }
\end{center}

In this work, we firstly investigate the trend of the measured SF half-lives for relatively stable nuclei. In Fig. 2(a), we show the measured maximum of $\log_{10} T_{\rm SF}$ so far in a certain isotopic chain as a function of neutron number. Simultaneously, we present in Fig. 2(b)  the corresponding relative barrier height $U_0=V_{\rm top}-Q_{\rm sf}+\Delta$ and the effective tunneling barrier $U$. Here, $V_{\rm top}$ and $Q_{\rm sf}$ denote the height of the Coulomb barrier between two fission fragments and the corresponding SF $Q$-value in symmetric fission, respectively. The Coulomb barrier is written as $V_{\rm top}=Z_1 Z_2 e^2/ (R_C^{(1)}+R_C^{(2)}+d)$ for symmetric fission, with the charge number $Z_1 \simeq  Z_2$ of the fission fragments. The corresponding charge radius $R_C^{(1)} \simeq R_C^{(2)}$ of the fission fragments at their ground state are given by the WS charge radius formula \cite{Wang13},
\begin{eqnarray}
	R_c =  R_0 \left [1+\frac{5}{8 \pi}(\beta_2^2 +
	\beta_4^2) \right ],
\end{eqnarray}
with which the 1014 measured charge radii can be reproduced with an rms error of only 0.021 fm \cite{Litao13}. In Eq.(4), the nuclear charge radius $R_0$ at spherical shapes is given by
\begin{eqnarray}
	R_0 = 1.226 A^{1/3} + 2.86 A^{-2/3} - 1.09 (I-I^2) + 0.99 \Delta E/A.
\end{eqnarray}
Here,  nuclear quadrupole deformation $\beta_2$, hexadecapole deformation $\beta_4$ and shell correction $\Delta E$ are taken from the WS4 model \cite{WS4}.  $I=(N-Z)/A$ denotes the isospin asymmetry. In the calculations of the Coulomb barrier $V_{\rm top}$, we introduce a separation distance $d$ between the fission fragments, considering that the fission fragments are in elongated shapes at the scission point. In addition, to consider the increase of the barrier height due to the effect of unpaired nucleons \cite{Bj80} (which will be discussed later) we adopt different values for the separation distance. $d=2.12$ fm for even-even nuclei and $d=1.83$ fm for nuclei with unpaired nucleons. The difference of the separation distance $d$ results in about $2.1\%$ change of the Coulomb barrier height for even-even actinides comparing to the neighboring odd-A nuclei, with $R_C^{(1)}+R_C^{(2)} \approx 12$ fm. We note that the average inner fission barrier heights \cite{Cap09} for odd-A actinides is higher than that of even-even actinides by about $2.5\%$, which indicates that adopting different separation distance $d$ is reasonable. 

$\Delta$ denotes the shell gap in the parent nucleus, which is to consider the influence of microscopic structure effects on the barrier and obtained from the difference of nuclear ground state energies \cite{Mo14},
\begin{eqnarray}
	\Delta (N,Z)=E(N+2,Z)+E(N-2,Z)+E(N,Z+2)+E(N,Z-2)-4E(N,Z).
\end{eqnarray}
It is known that in addition to barrier height, the barrier width can also affect the fission probability according to WKB calculations. From the degree-of-freedom of elongation, the fission barriers of nuclei with spherical shapes could be thicker than those with prolate deformations. The shell gaps in doubly-magic nuclei (with  spherical shapes) are generally larger than those in mid-shell nuclei (with prolate deformations).  To effectively consider the influence of barrier width, we therefore add the shell gap $\Delta$ in the calculations of the relative Coulomb barrier height $U_0$. 

The effective tunneling barrier (ETB) is defined as $U=U_0-Q_\beta/x$, in which the influence of isospin effects on the barrier height is considered for nuclei far from the $\beta$-stability line. Here, $Q_{\beta}$ denotes the total $\beta$-decay energy for a nucleus, which is obtained by calculating the difference between the ground state energy of the parent nucleus and the corresponding energy of the $\beta$-stability nucleus (with the same mass number). $x=\frac{E_c}{2E_s}$ denotes the fissionability parameter \cite{Co63}, with the Coulomb energy $E_c=a_c Z^2/A^{1/3}$ and the surface energy $E_s=a_s A^{2/3} (1- \kappa I^2)$ of the nuclei in which the coefficients are taken from WS4 \cite{WS4}.

\begin{figure}
	\setlength{\abovecaptionskip}{ -1.0 cm}
	\includegraphics[angle=0,width=0.85\textwidth]{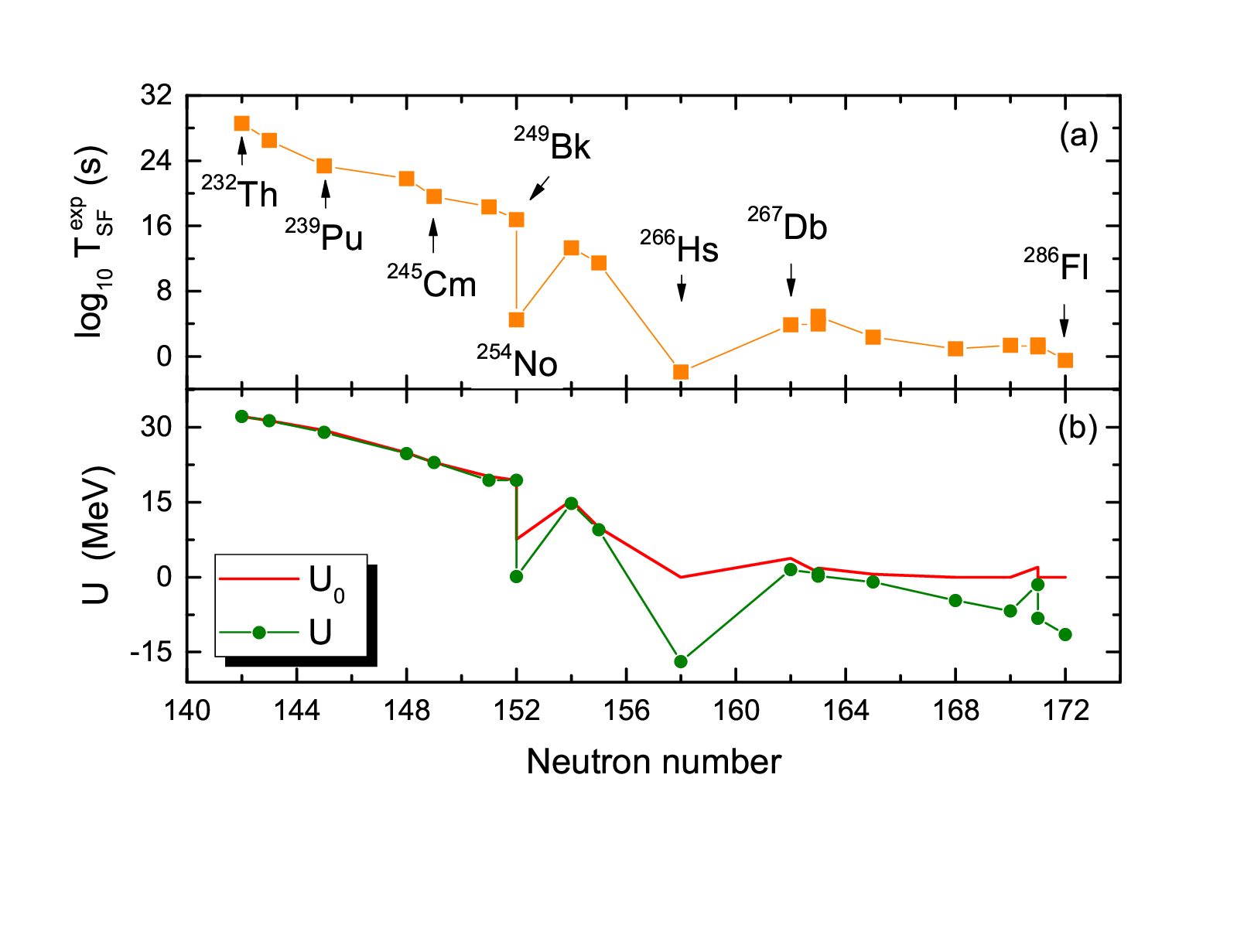}
	\caption{(Color online)  (a) Measured maximum of  $ \log_{10} T_{\rm SF}$ in each isotopic chain with mass numbers from $A=232$ to $A=286$.  (b) The corresponding values of the effective tunneling barrier $U$. The red curve denotes the relative barrier height $U_0=V_{\rm top}-Q_{\rm sf}+\Delta$.  }
\end{figure}

From Fig. 2, one can see that the trend of the effective tunneling barrier $U$ are in good agreement with that of $\log_{10} T_{\rm SF}$, including the abrupt falls of the half-lives for $^{254}$No and $^{266}$Hs, which implies that the effective tunneling barrier plays a role for the SF half-lives. Comparing $U$ with $U_0$, one  notes that the abrupt falls of the half-lives for neutron-deficient $^{254}$No and $^{266}$Hs could be due to the $\beta$-decay energy $Q_\beta$, which provides an additional energy to overcome the barrier like in the process of $\beta$-delayed fission.  Considering the possible correlations between ETB and  $\log_{10} T_{\rm SF}$, the SF half-lives for heavy and super-heavy nuclei are expressed as,
\begin{eqnarray}
	\log_{10}  T_{\rm SF} (s) = c_1 U+ c_2 U^2 + \Delta_{\rm res}.  
\end{eqnarray}
The term $\Delta_{\rm res}$ is introduced to consider the influence of the pairing effects and residual correlations, 
\begin{eqnarray}
	\Delta_{\rm res}= \left\{
	\begin{array} {l@{\quad:\quad}l}
		\,\,\, \Delta + \sin(c_3 Q_\alpha)  &   N {\rm ~and~} Z {\rm ~even }    \\
	    \frac{3}{2}\Delta + \sin(c_3 Q_\alpha + \frac{1}{2}\Delta)  &    A {\rm ~ odd } \\
		2 \Delta + \sin(c_3 Q_\alpha + \frac{1}{2}\Delta)   &   N {\rm ~and~} Z {\rm ~ odd } .   \\
			\end{array} \right.
\end{eqnarray}
Where, the $Q_\alpha$ term is an empirical correction to consider the correlation between $\log_{10} T_{\rm SF}$ and $Q_\alpha$  \cite{Anto25}, which will be discussed later. By fitting the measured SF half-lives for 79 known nuclei with certain branching ratios in NUBASE2020 \cite{NUBASE2020}, the optimal model parameters $c_1=0.466$, $c_2=0.0088$, $c_3=1.98$ are obtained. 

For nuclei with odd proton and/or odd neutron numbers, the change of the energy of the fissioning state along the fission path plays a decisive role. Due to spin and parity conservation at crossing points of Nilsson levels, the unpaired nucleon in general cannot change the level as it is the case for nucleon pairs in even-even nuclei, which leads to an effective increase of the fission barrier \cite{Hess}. It is therefore found that the half-lives of odd-mass nuclei are systematically longer than its neighboring even-even nuclei due to the unpaired nucleons \cite{Rod16}. In this work, different values of $\Delta_{\rm res}$ are adopted to consider the pairing effects.

For even-even parent nuclei, the unpaired nucleons of fission fragments could lead to a relatively higher barrier in symmetric fission due to the lower $Q$-value. To consider the influence of the unpaired nucleons, in the calculations of $U_0$ we take the mean value of the $Q_{\rm sf}$ among three cases: with completely symmetric fragments, with one more neutron for a fragment, and with one more proton for a fragment. As an example, the three channels  $^{254}$Fm $\rightarrow ^{127}$Sn+$^{127}$Sn, $^{254}$Fm $\rightarrow ^{128}$Sn+$^{126}$Sn and $^{254}$Fm $\rightarrow ^{127}$Sb+$^{127}$In are considered in the calculations of $U_0$ for $^{254}$Fm. We note that taking the mean value for $Q_{\rm sf}$ can obviously improve the model accuracy for even-even nuclei. In addition, in the calculations of $U_0$ we introduce a truncation, i.e. $U_0 \geqslant 0 $, to avoid negative barrier height. 

\begin{center}
\textbf{ III. RESULTS AND DISCUSSIONS }
\end{center}

\begin{figure}
	\setlength{\abovecaptionskip}{ -0 cm}
	\includegraphics[angle=0,width= 0.75\textwidth]{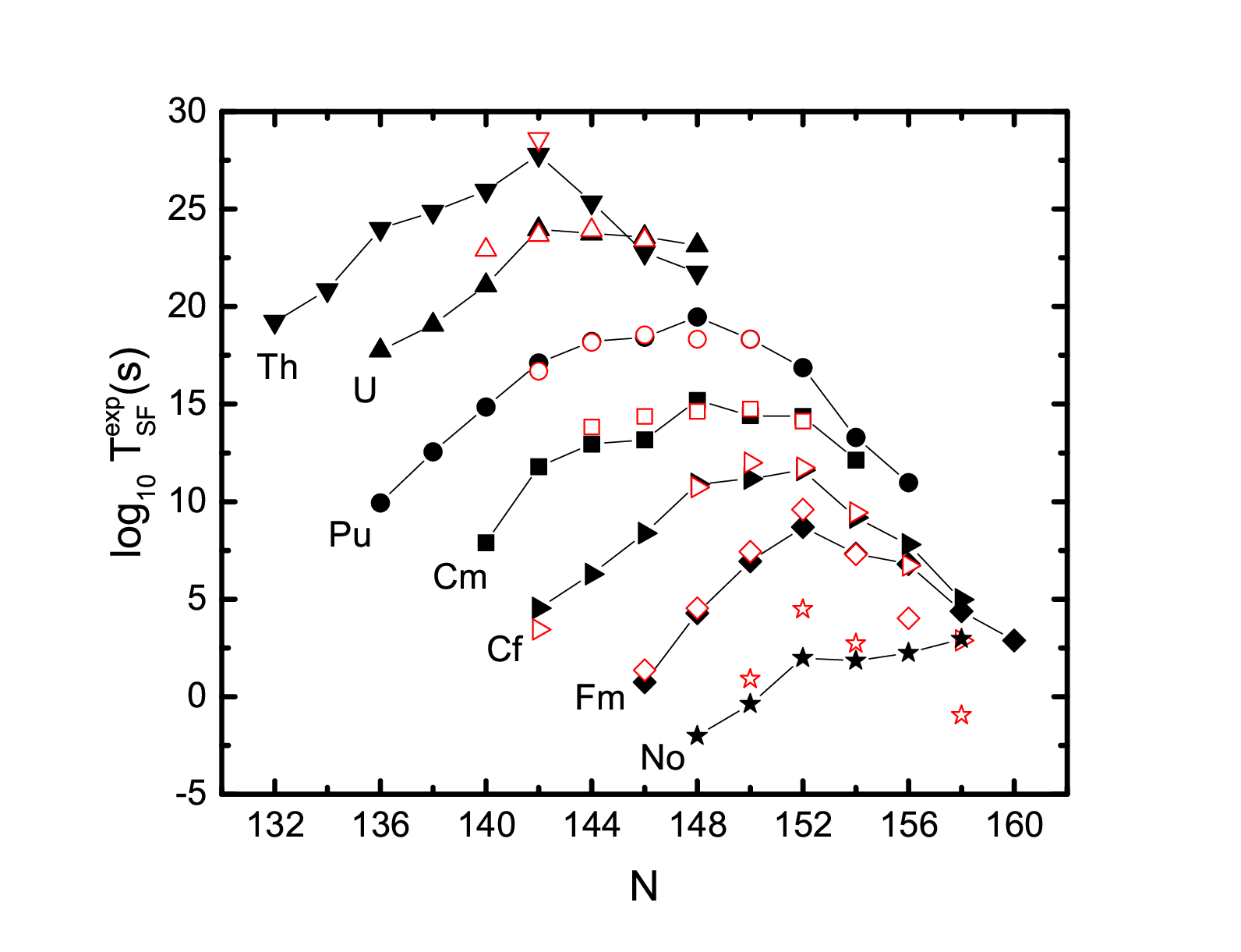}
	\caption{(Color online) Spontaneous fission half-lives of even-even actinides. The black curves denote the calculated results with Eq.(7) and the red open symbols denote the experimental data taken from NUBASE2020 \cite{NUBASE2020}.   }
\end{figure}

\begin{figure}
	\setlength{\abovecaptionskip}{ -2 cm}
	\includegraphics[angle=0,width= 0.75\textwidth]{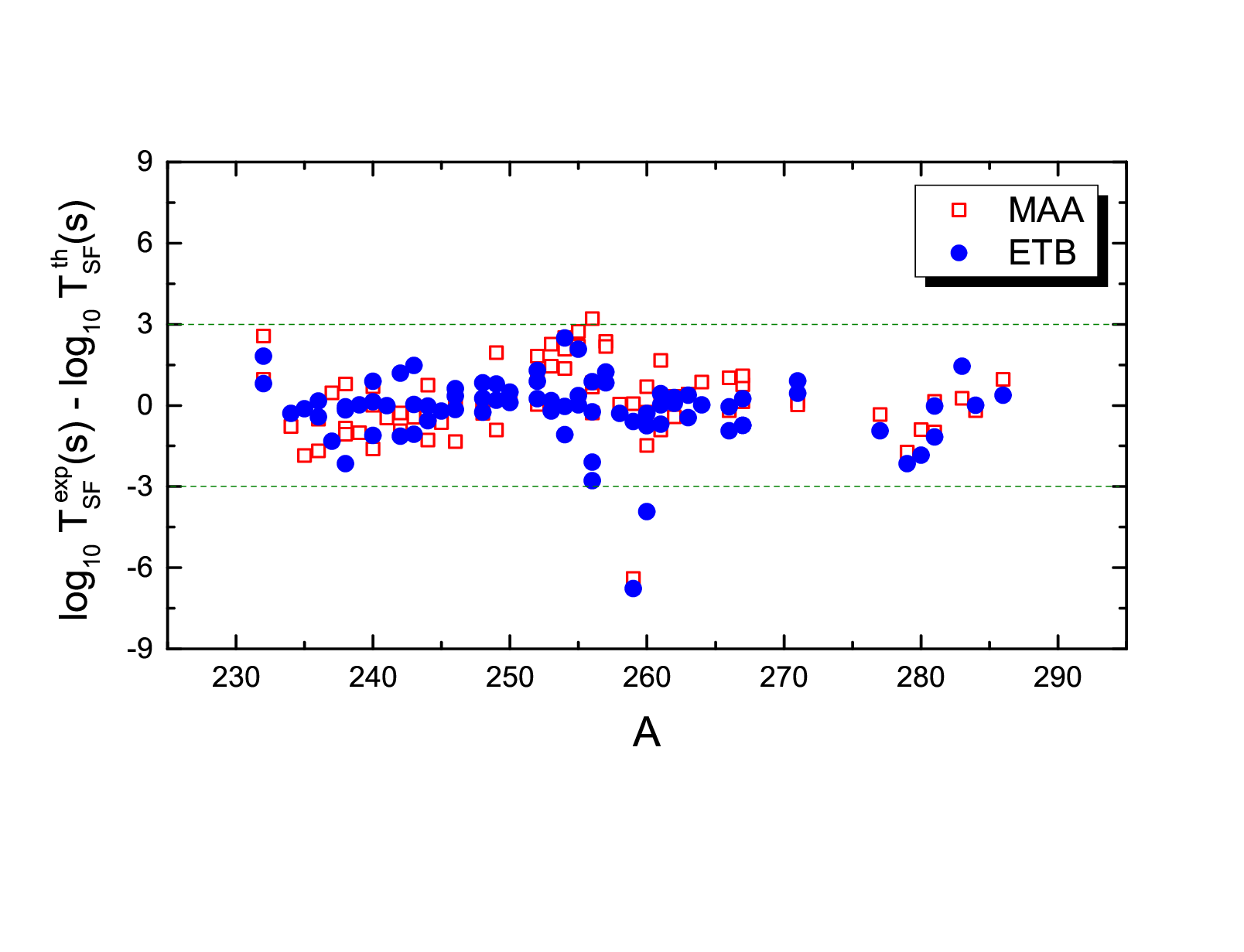}
	\caption{(Color online) Deviations of model predictions from the experimental half-lives. The circles and the squares denote the results with the proposed ETB approach and those of the MAA formula \cite{Anto25}, respectively.  }
\end{figure}

In Fig. 3, we show the predicted spontaneous fission half-lives of even-even actinides as a function of neutron numbers.  The curves denote the calculated results with Eq.(7). Open symbols denote the experimental data taken from NUBASE2020.  From Fig. 3, one can see that the measured SF half-lives can be reproduced reasonably well. In addition, we also note that the SF half-lives of neutron-deficient Cf and Fm isotopes are evidently shorter than those of the corresponding nuclei around the $\beta$-stability line, which also indicates that the $\beta$-decay energy $Q_\beta$ may play a role in the decreasing behavior of SF half-lives for nuclei far from the $\beta$-stability line.

Fig. 4 shows the deviations between the measured SF half-lives and model predictions. The circles and the squares denote the results with the proposed ETB approach and those of MAA formula \cite{Anto25}, respectively. One sees that for almost all nuclei, the deviations between the calculated half-lives and experimental data are within three orders of magnitude. By using Eq.(9), the average deviation between model predictions and experimental data 
\begin{eqnarray}
\langle\sigma\rangle = \frac{1}{n} \sum_{i=1}^{n} \left| \log_{10}   (T_{\rm SF}^{{\rm th}, i} /T_{\rm SF}^{{\rm exp}, i})  \right|
\end{eqnarray}
 is only 0.800 for the 79 known nuclei. For $^{259}$Fm, we note that its SF half-life is significantly over-predicted by both MAA and ETB calculations.  
In Table I, we list the calculated average deviations $\langle\sigma\rangle $ with respect to the SF half-lives of 79 nuclei (including 42 even-even nuclei), by using two phenomenological models (MAA and ETB). In the calculations, the $\alpha$-decay energy $Q_\alpha$ are taken from WS4 plus radial basis function (WS4+RBF) predictions \cite{WS4}, with which the known $\alpha$-decay energies of super-heavy nuclei can be reproduced with an rms error of 0.22 MeV \cite{ Wang15}. For 42 even-even nuclei, the average deviation is $\langle\sigma\rangle=0.783$ with the proposed ETB approach, which is smaller than that with XRG formula \cite{Ren08} and with MAA formula \cite{Anto25} by $28\%$ and $13\%$, respectively. It should be mentioned that in \cite{Anto25} the SF half-lives of 111 nuclei (including 58 even-even nuclei) are used for analysis. 

\begin{figure}
	\setlength{\abovecaptionskip}{ -0 cm}
	\includegraphics[angle=0,width= 0.75\textwidth]{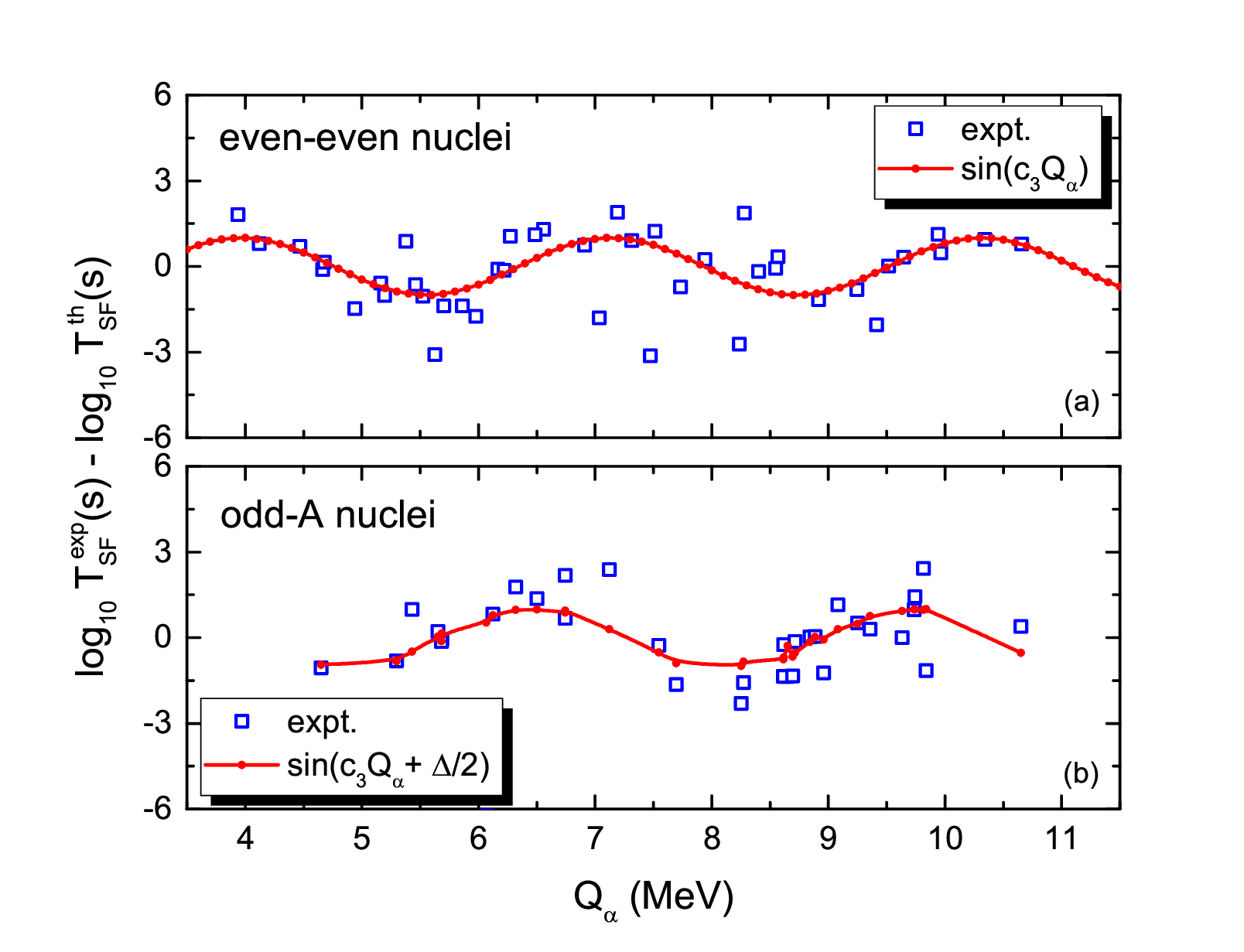}
	\caption{(Color online) Deviations between the measured half-lives and the predictions with Eq.(7) but neglecting the $Q_\alpha$ term. (a) for even-even nuclei and (b) for odd-A nuclei. }
\end{figure}

\begin{table}    	
	
	\caption{ Average deviation between model predictions and 79 experimental SF half-lives taken from NUBASE2020 \cite{NUBASE2020}.  }
	\begin{tabular}{ccc}
		\hline\hline
		
		~~~model~~~  & ~~~even-even nuclei~~~ &~~~ all nuclei~~~    \\
		\hline
		MAA     &  0.900    &  0.959       \\
		ETB     &  0.783    &  0.800       \\
				
		\hline\hline
	\end{tabular}
\end{table}

  To investigate the residual correlations between $\log_{10} T_{\rm SF}$ and $Q_\alpha$, we show in Fig. 5 the deviations between the measured half-lives and the predictions with the ETB approach but neglecting the $Q_\alpha$ term in the calculations. It seems that there exists somewhat oscillatory behavior in the deviations. In Eq.(8), we adopt a sine function to describe the correlations between $\log_{10} T_{\rm SF}$ and $Q_\alpha$ rather than a linear relationship, considering that many properties of atomic nuclei (such as binding energy and fission barrier) exhibit oscillatory behavior similar to a sine wave with changes in nucleon number, particularly when crossing shell closures. The shell correction energy itself represents an oscillation around the smooth values predicted by the liquid-drop model.  The sine function captures the oscillatory shell corrections, similar to its use in modeling the residual shell effects in nuclear masses \cite{Wang10}. In addition, comparing to the linear correlation, the sine function is a bounded function with a range of $[-1, 1]$. This means that no matter how large $Q_\alpha$ becomes, the contribution of the correlation term always has an upper and lower limit, indicating the existence of a saturation mechanism with which the unphysical and infinite contribution can be avoided. With the phenomenological sine functions to describe the oscillatory behavior, the average deviation $\langle\sigma\rangle$ is significantly reduced by about $20\%$.

 \begin{figure}
	\setlength{\abovecaptionskip}{ -0.  cm}
	\includegraphics[angle=0,width= 0.95\textwidth]{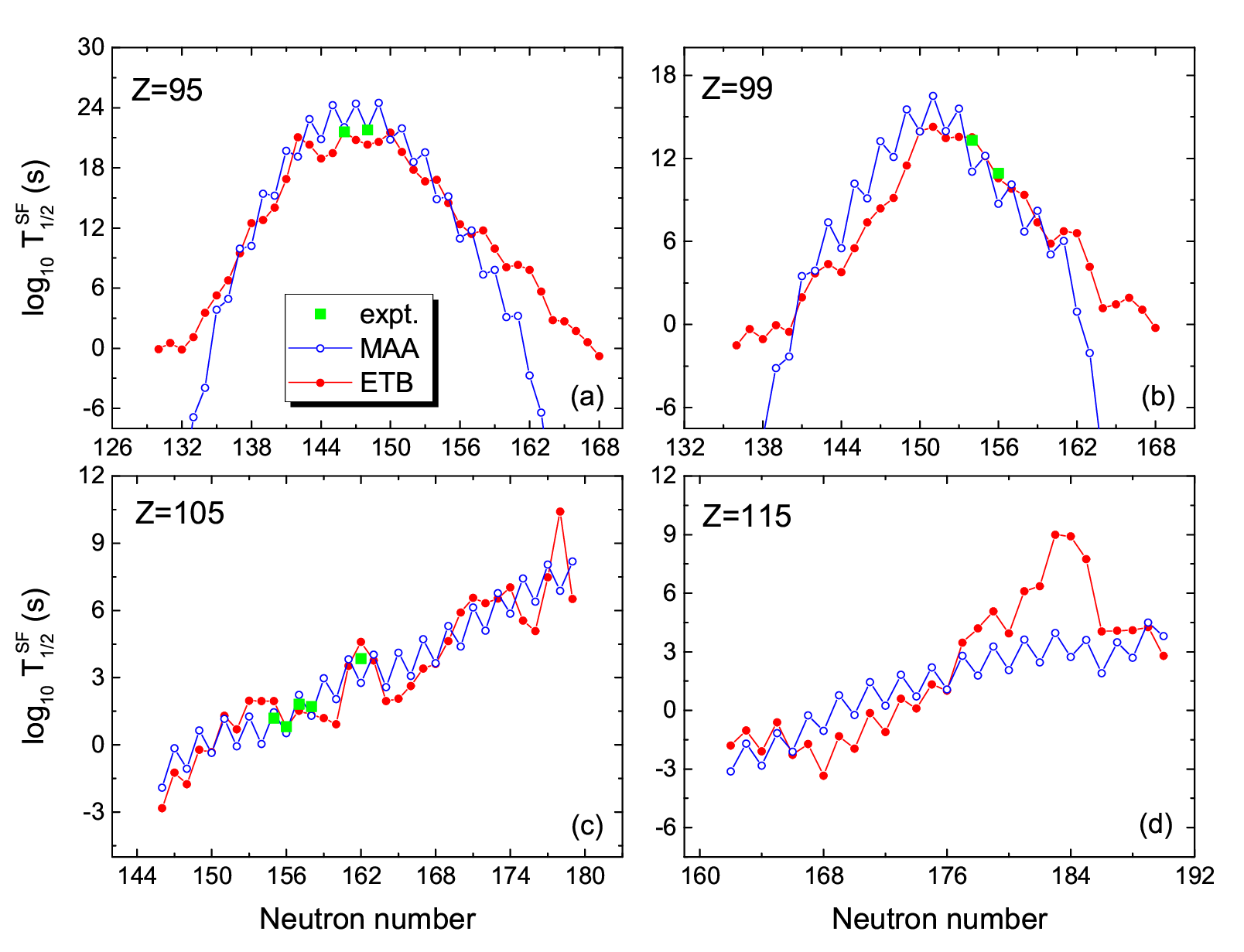}
	\caption{(Color online) SF half-lives for Am, Es, Db and Mc isotopes. The red curves with solid circles denote the predictions with the ETB approach proposed in this work. The blue curves with open circles denote the results of MAA \cite{Anto25}. The squares denote the experimental data taken from NUBASE2020 \cite{NUBASE2020}.}
\end{figure}

   In Fig. 6, we show the predicted spontaneous fission half-lives for Am, Es, Db and Mc isotopes. The squares denote the data taken from NUBASE2020. One sees that the SF half-lives can be reproduced well with both models for known nuclei. For Am and Es isotopes in the range of $40<N-Z<62$, the predicted results from the two models are in good agreement with each other generally. For Am and Es isotopes far from the $\beta$-stability line the predicted SF half-lives with MAA formula are significantly smaller than those with Eq.(7). For some extremely neutron-deficient nuclei such as $^{214}$U \cite{Zhang21}, we note that the predicted SF half-lives from the MAA formula are unphysical and catastrophically short (see Fig. 1). In MAA model, the authors emphasized that extrapolation to nuclei well outside the range $40<N-Z<62$ should be treated with caution, since the linear correlation used in the formulas is observed from the available experimental SF half-lives in the following ranges: $90 \leqslant Z \leqslant 102$ and $41 \leqslant N-Z \leqslant 60$, as well as $103 \leqslant Z \leqslant 118$ and $45 \leqslant N-Z \leqslant 61$. In the ETB calculations, we introduce a truncation $Q_{\beta} \leqslant  a_{\rm sym}$ for nuclei around driplines, with the symmetry energy coefficient  $a_{\rm sym} = c_{\rm sym}\left [1-\frac{\kappa}{A^{1/3}}+ \xi  \frac{2-|I|}{2+|I|A} \right ] $ taken from WS4.  $a_{\rm sym}$ represents the upper energy limit of neutron-proton asymmetry that a nuclear system can sustain. The truncation for $Q_{\beta}$ ensures that the model's predictions, even under extreme conditions, do not deviate entirely from physical reality. For Db isotopes, the results from the two models are quite close to each other and the peaks at $N=162$ and $N=178$ are due to the shell effects in the calculations of ETB. For Mc isotopes around neutron magic number $N=184$, the predicted SF half-lives with the proposed method are significantly enhanced due to the strong shell effects.

 \begin{figure}
	\setlength{\abovecaptionskip}{ 0 cm}
	\includegraphics[angle=0,width= 0.95 \textwidth]{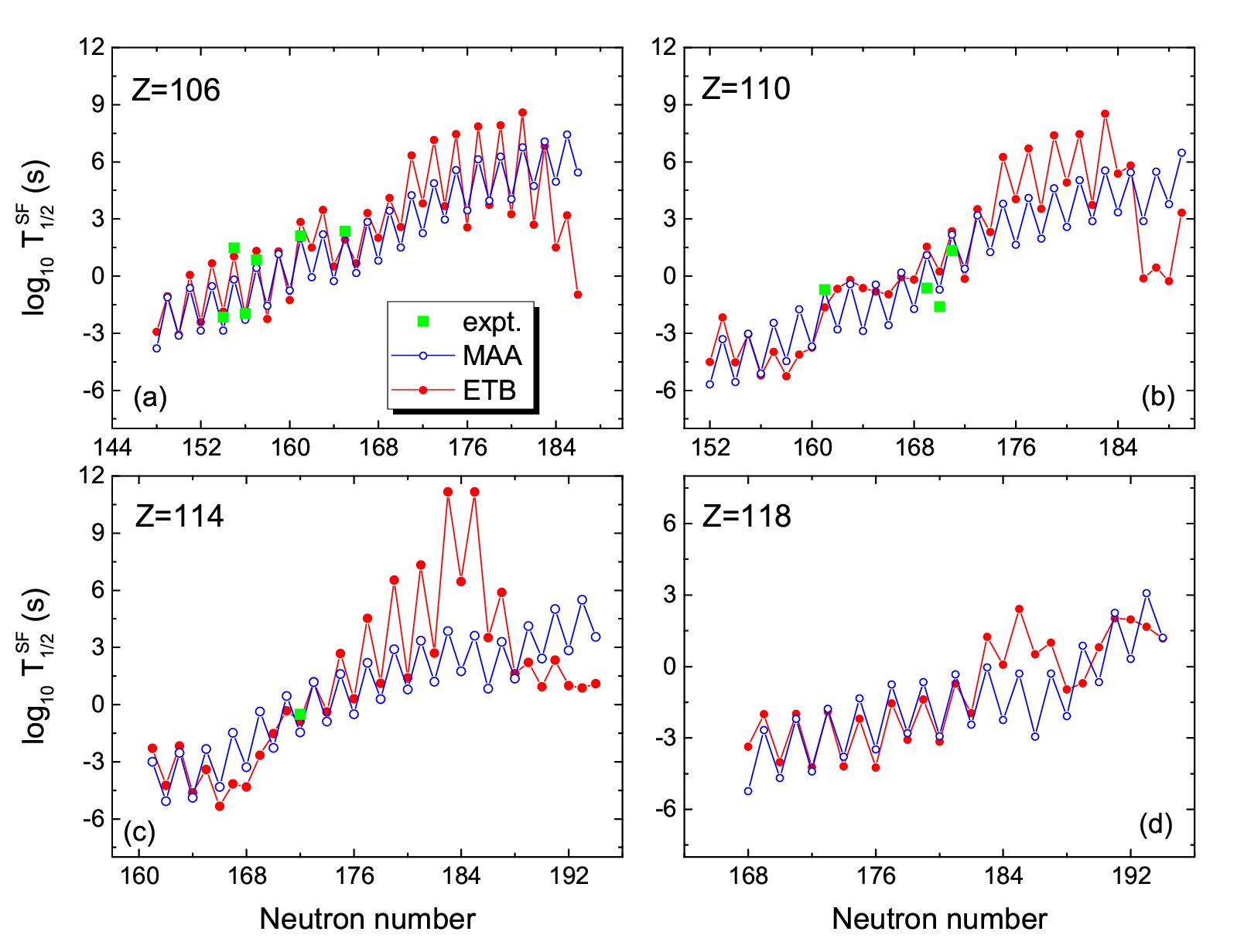}
	\caption{(Color online) The same as Fig. 6, but for Sg, Ds, Fl and Og isotopes.}
\end{figure}

Fig. 7 shows the predicted spontaneous fission half-lives for Sg, Ds, Fl and Og isotopes. We note that the available experimental data from NUBASE2020 can be well reproduced by these two phenomenological models. Simultaneously, the odd-even staggering due to the pairing effects can be clearly observed from the results of both models. For Fl and Og isotopes, the predicted SF half-lives around $N=184$ with ETB are significantly larger than those with MAA formula. In the region of $45 \leqslant N-Z \leqslant 61$, the predicted results from these two models are in good agreement with each other. From Fig. 7 and Fig. 8 in Ref. \cite{Mo14}, we note that the uncertainties of the predicted shell gaps from different mass models are still large for super-heavy nuclei. The obtained shell gaps from WS4 are $\Delta=3.40$ MeV for $^{272}$Ds, 5.58 MeV for $^{298}$Fl, and 3.76 MeV for $^{304}120$. The pronounced enhancement of the SF half-lives for nuclei around $^{298}$Fl is due to the large shell gaps, since $\log_{10} T_{\rm SF}$ is directly related to $\Delta$ according to Eq.(8).

 \begin{figure}
	\setlength{\abovecaptionskip}{ -4 cm}
	\includegraphics[angle=0,width= 0.95 \textwidth]{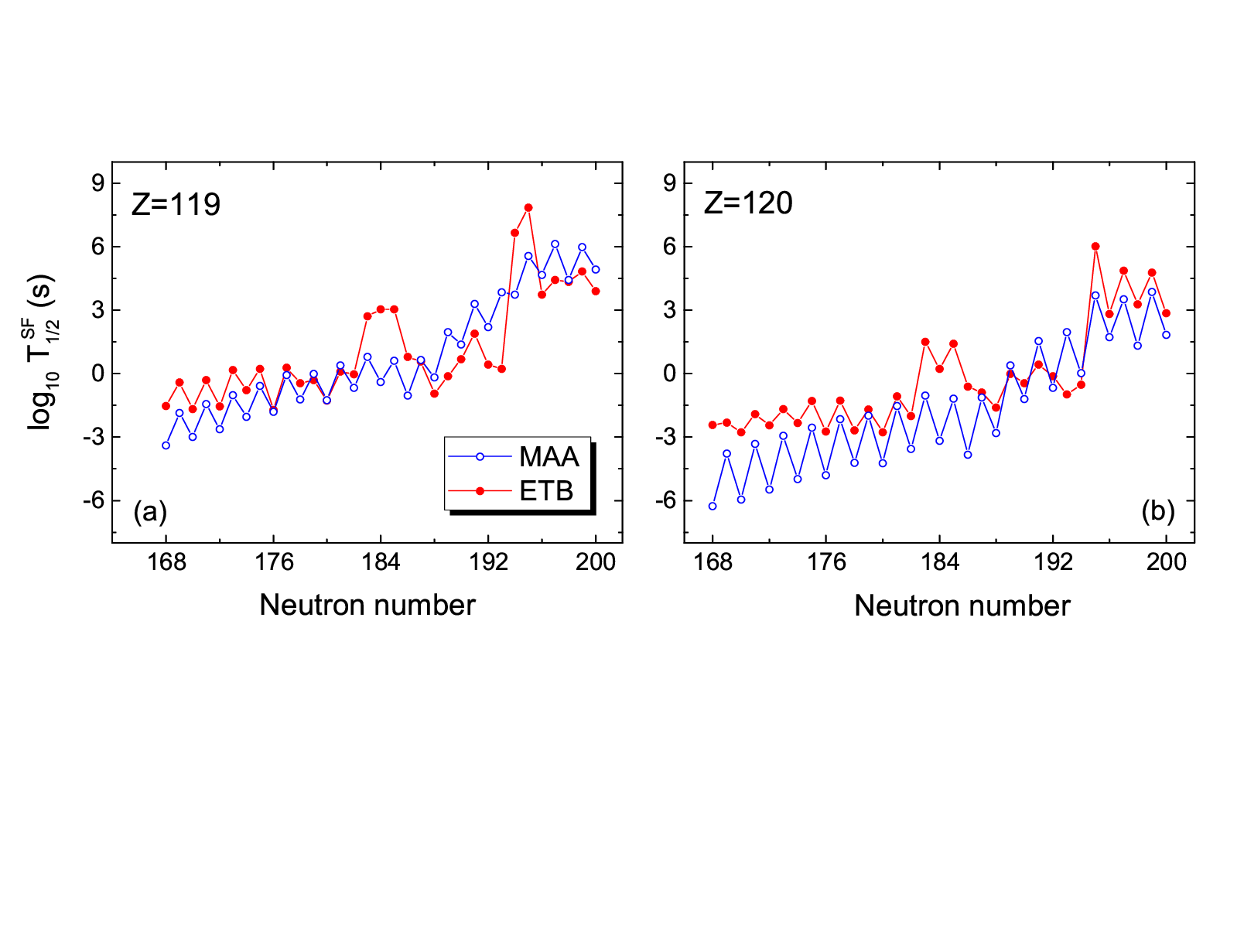}
	\caption{(Color online) The same as Fig. 6, but for nuclei with $Z=119$ and $Z=120$.}
\end{figure}

In Fig. 8, we present the predicted SF half-lives for nuclei with $Z=119$ and $Z=120$. The trends of the predicted SF half-lives for neutron-rich nuclei from the two model are close to each other. At neutron-deficient side, the results of ETB approach proposed in this work are slightly higher than those of MAA. For SHN $^{293}119_{174}$, the predicted value of $T_{\rm SF}$ is about $10 \sim 160$ milliseconds based on MAA and ETB calculations.  Considering that the average deviations between data and model predictions from the two models are within one order of magnitude, the SHN $^{293}119$ that might be synthesized in fusion reaction  $^{54}$Cr+$^{243}$Am after evaporating four neutrons, can survive for much longer than $\sim$1 $\mu$s to reach the focal-plane detector.

\begin{center}
\textbf{IV. SUMMARY}
\end{center}

 In this study, we compared two phenomenological models for systematically describing the spontaneous fission (SF) half-lives $T_{\rm SF}$  of heavy and superheavy nuclei. Based on the effective tunneling barrier (ETB) which considers the relative Coulomb barrier between fission fragments, the shell gap $\Delta$ and  $\beta$-decay energy $Q_\beta$ of the fissioning nuclei, the SF half-lives of 79 known nuclei can be reproduced with an average deviation of 0.8, which is smaller than those of two other phenomenological models: XRG and MAA formulas. With a value of $U=28.8$ MeV, the height of ETB for $^{236}$U is higher than that of $^{246}$Cm ($U=19.5$ MeV) by 9.3 MeV, which explains why the SF half-life of $^{236}$U is up to 9 orders of magnitude longer than that of $^{246}$Cm. The competition between Coulomb and isospin effects, represented by the $\beta$-decay energy $Q_\beta$, significantly impacts SF half-lives of nuclei far from the $\beta$-stability line. For superheavy nuclei with $45 \leqslant N-Z \leqslant 61$, the predicted SF half-lives from two different phenomenological models (ETB and MAA) are in good agreement with each other generally. For nuclei around neutron magic numbers such as $N=162$ and $N=184$, the predicted SF half-lives from ETB are larger than those from MAA since the shell gaps are directly involved in the ETB calculations. Both the proposed ETB approach and the MAA formula predict remarkably shorter SF half-life for superheavy nucleus $^{304}$120 comparing with $^{298}$Fl. For nuclei with $Z=119$ and $N=174$, the predicted $T_{\rm SF}$ from ETB and MAA is about $10 \sim 160$ ms, which is longer than the corresponding $\alpha$-decay half-life \cite{Tian24} by two or three orders of magnitude. These results provide helpful insights for future experiments aimed at synthesizing new superheavy elements and understanding their stability.

\begin{center}
\textbf{ACKNOWLEDGEMENTS}
\end{center}
This work was supported by National  Key R\&D Program of China (No. 2023YFA1606503), National Natural Science Foundation of China (Nos. 12265006, 12035001,U1867212)  and Guangxi "Bagui Scholar" Teams for Innovation and Research Project. N. W. is grateful to Dr. Zhiyuan Zhang for valuable discussions. The predicted SF half-lives for 1311 nuclei are available on http://www.imqmd.com/decay/

\end{document}